\documentclass[11pt]{article}%
\usepackage{amsmath}
\usepackage{subeqn}
\usepackage{amsfonts}%
\usepackage{amssymb}
\newcommand{\ben}{\begin{enumerate}}
\newcommand{\een}{\end{enumerate}}
\newcommand{\be}{\begin{equation}}
\newcommand{\bse}{\begin{subequation}}
\newcommand{\ese}{\end{subequation}}
\newcommand{\bea}{\begin{eqnarray}}
\newcommand{\eea}{\end{eqnarray}}
\newcommand{\bc}{\begin{center}}
\newcommand{\ec}{\end{center}}

\def\eqn#1{\be\label{#1}}
\def\ce{{\cal E}}
\def\bea{\begin{eqnarray}}
\def\eqnn#1{\bea\label{#1}}
\def\eea{\end{eqnarray}}

\def\hn{\hat{n}}

\def\half{{\textstyle{\frac{1}{2}}}}

\newcommand{\eqna}[1]{\begin{subequations} \label{#1}
\begin{eqnarray}}
\def\eena{\end{eqnarray}
\end{subequations}}
\begin{document}

\begin{center}

{\Large\bf Revisiting the Quantum Group Symmetry of Diatomic Molecules}

 \vspace{7mm}

{\textbf{Maia Angelova}$^{\dagger }$\textbf{, Vladimir
Dobrev}$^{\dagger }$\footnote{Permanent address:
Institute of Nuclear Research and Nuclear Energy,
Bulgarian Academy of Sciences, Sofia, Bulgaria.}
\textbf{\ and Alejandro Frank}$^{\ddagger }$

\vspace{5mm}

$^{\dagger }$\textit{School of Informatics}\\
\textit{Northumbria University, Newcastle-upon-Tyne NE1 8ST, UK}\\[2mm]
$^{{}}\ddagger $\textit{Instituto de Ciencias Nucleares and Centro de
Ciencias F\'{i}sicas,}\\
\textit{UNAM, A.P. 70-543, Mexico, D.F., 04510 Mexico.}\\
}

\end{center}

\vspace{.8 cm}

\begin{abstract}
We propose a $q$-deformed model of the anharmonic vibrations in
diatomic molecules. We analyse the applicability of the model to
the phenomenological Dunham's expansion by comparing with
experimental data. Our methodology involves a global consistency
analysis of the parameters that determine
the $q$-deformed system, when compared with fitted
vibrational parameters to 161 electronic states in diatomic
molecules. We show how to include both the positive and the
negative anharmonicities in a simple and systematic fashion.
\end{abstract}

PACS: 33.15.Mt, 02.20.Uw, 31.15.Hz, 03.65.Fd, 02.20.-a

\vspace{1 cm}

\section{Introduction}

$Q$-algebras were originally conceived as a means to solve the
quantum Yang-Baxter equation \cite{DrJi}, but in the last decade
a great number of applications have been found in diverse areas
of physics, ranging from the deformation of conformal field
theories \cite{AGGS} to optics \cite{BuCh} and nuclear and
molecular spectroscopy \cite{BADRRS},\cite{RRS},\cite{BRF}. On
the other hand, algebraic models have been proposed and applied
systematically in many fields, including in a relevant way the
study of nuclei and molecules \cite{ArIa},\cite{FrVI}. This kind
of approach has simplified the n-body problem dramatically and
has given rise to numerous new insights, including, for example,
a supersymmetric description of quartets of nuclei and the
formulation of tractable models of polyatomic molecules
\cite{Fra}. In the latter case, these methods combine the use of
Lie algebraic and discrete symmetry techniques, which describe
the interactions and the global symmetry, respectively, of these
systems. In particular, the $su(2)$ algebra has been proposed as a
basic algebraic structure, given its connection to the Morse
potential \cite{FrVI}. This allows an algebraic treatment of
anharmonicities, which become increasingly important as higher
vibrational modes are excited, much in the same fashion as the
harmonic oscillator algebra is used in connection with harmonic
motion.

While the Morse potential leads to a fairly good approximation to
the spectroscopic properties of most diatomic molecules, it has
some limitations. For example, its energy eigenvalues contain
only quadratic corrections to the equally-spaced harmonic
behavior, while additional, higher-order terms are usually
required in the phenomenological Dunham expansion to produce an
accurate fit to the observed vibrational energies.

In itself, the Dunham expansion does not constitute an adequate
theoretical framework, since no Hamiltonian, and hence no
eigenfunctions, can be deduced from it. It is thus an interesting
problem to see whether an algebraic framework can be constructed
in such a way as to include these higher order corrections, while
maintaining its simplicity and providing a Hamiltonian and
corresponding wave functions.

In this paper we report on an investigation to extend the $su(2)$
framework, by considering the $q$-deformation of the Morse
Hamiltonian. We then test our ideas by comparing a series
expansion of the $q$-Morse energy expression with experimental
Dunham expansions for a large number of molecules. As a first
step we restrict our analysis to test the consistency of this
$q$-extension with respect to the Dunham expansion. A subsequent
and more delicate test should include the evaluation of the
$q$-deformed Morse vibrational eigenfunctions and the calculation
of dipole intensities in the same fashion as for the $su(2)$ model.
We wish to emphasise that, while other studies have considered
the deformation of both vibrational and rotational degrees of
freedom in molecules \cite{BADRRS},\cite{RRS},\cite{BRF}, our
approach is simpler and more basic, as we aim to generalise an
established model for vibrational spectroscopy, that is not
restricted to diatomic molecules but can be extended to
polyatomic systems, by $q$-deforming the fundamental bond
interaction. In addition, we test the $q$-deformation in a global
way by comparing with the Dunham parameters of 161 electronic
states of diatomic molecules.

We start the next section by discussing the traditional treatment
of anharmonic vibrations via the Dunham expansion.

\section{Molecular Anharmonic Vibrations}

The phenomenological description of the vibrational energy of
diatomic molecules in a given electronic state is given by the
Dunham's expansion \cite{dunham:1932}, traditionally written as:
\begin{equation}
E(v)=hc\omega_{e}\left( v+\frac{1}{2}\right) -hc\omega_{e}x_{e}\left(
v+\frac{1}{2}\right) ^{2}+hc\omega_{e}y_{e}\left( v+\frac{1}{2}\right)
^{3}+\ldots\;. \label{dunham}%
\end{equation}
where $c$ is the speed of light in vacuum and $v$ is the
vibrational quantum number. The vibrational molecular constants
$\omega_{e}$, $\omega_{e}x_{e}$ and $\omega_{e}y_{e}$ are
obtained by fitting the potential curve to the experimental
spectral data, $\omega_{e}x_{e}\ll\omega_{e}$,\ \ $\omega
_{e}y_{e}\ll\omega_{e}x_{e}$. While the constant
$\omega_{e}x_{e}$ is nearly always positive, the constant
$\omega_{e}y_{e}$ can be positive or negative and is often very
small \cite{herz:dia}. The terms of quadratic and higher order
with respect to $(v+\frac{1}{2})$ in the expansion (\ref{dunham})
account for the anharmonic character of molecular vibrations and
the constant $x_{e}$ is called the anharmonicity constant.

The expansion (\ref{dunham}) is part of the more general Dunham's
description of the rotational-vibrational energy of molecules
\cite{{dunham:1932},{OT}}
\begin{equation}
E(v,j)=\sum_{l,m}y_{lm}\left( v+\frac{1}{2}\right) ^{l}\left(
J(J+1)\right) ^{m} \label{gdunham}
\end{equation}
where $J$ is the angular momentum and the coefficients $y_{lm}$
are often called Dunham coefficients. \ Here,
$y_{10}=hc\omega_{e}$,\ $y_{20} =-hc\omega_{e}x_{e}$,
$y_{30}=hc\omega_{e}y_{e}$, and the vibrational part is obtained
by taking the terms with $m=0$, i.e. by ignoring the rotational
bands built on each vibrational bandhead. These parameters are
essential since structural information is contained fundamentally
in the vibrational spectra. Rotation-vibration interaction can
usually be treated in a perturbative fashion.

Dunham's expansion provides a convenient, empirical and model-free
way of organising a large quantity of spectral data. It also
provides a procedure for comparing data with the calculations
arising from model potentials such as the Morse potential. This
expansion, however, has the disadvantage that it does not arise
from a Hamiltonian and hence does not provide wave functions for
the vibrational states.

If the expansion (\ref{dunham}) is truncated to the quadratic
term, one obtains essentially the discrete spectrum of the Morse
potential
\begin{equation}
E_{M}(v)=hc\omega_{e}\left( \left( v+\frac{1}{2}\right) -x_{e}\left(
v+\frac{1}{2}\right) ^{2}\right) . \label{morse}
\end{equation}
The Morse potential \cite{morse1929},
\begin{equation}
V_{M}(r-r_{e})=D_{e}\left( 1-e^{-\beta(r-r_{e})}\right) ^{2}
\end{equation}
describes reasonably well the potential energy of diatomic
molecules near the equilibrium position $r_{e}$ and accounts for
the anharmonicity of the molecular vibrations. It is
characterised by two parameters, $D_{e}$ is the depth of the
minimum of the curve and $\beta$ (known as restitution constant
in the harmonic approximation's models), $\beta>0$. Using the
Hamiltonian,
\begin{equation}
H_{M}=-\frac{\hbar}{2\mu}\nabla^{2}+V_{M}
\end{equation}
Morse \cite{morse1929} solved exactly the Schr\"{o}dinger
equation, and found the quantised energy levels
\begin{equation}
E_{M}(v)=hc\left( \beta\sqrt{\frac{D_{e}\hbar}{\pi c\mu}}\left(
v+\frac{1}{2}\right) -\frac{\hbar\beta^{2}}{4\pi c\mu}\left( v+\frac{1}
{2}\right) ^{2}\right) . \label{morse2}
\end{equation}
Here $\mu$ is the reduced mass of the molecule. Comparing the
coefficients in the expressions (\ref{morse}) and (\ref{morse2}),
the well-known relations between the Morse parameters $D_{e}$ and
$\beta$ and the Dunham coefficients are obtained
\begin{equation}
D_{e}=\frac{\omega_{e}}{4x_{e}}\text{ \ \ and \ \
}\beta=\sqrt{\frac{4\pi
c}{\hbar}\mu\omega_{e}x_{e}}.
\end{equation}
These relations are also known as consistency conditions \cite{herz:dia},
\cite{atkins:mol}.

One significant feature of the Morse potential is that the number
of bound energy levels is finite, i.e. $v=0,1,2,\ldots\left[
v_{M}\right] $, where $\left[ v_{M}\right]$ denotes the largest
integer less or equal to $v_{M}$. Indeed, the Morse energy
(\ref{morse}) has a maximum at $v_{M}$,
\begin{equation}
v_{M}=\frac{1}{2}\left( \frac{1}{x_{e}}-1\right) . \label{mlevel}
\end{equation}
and thus the number of levels in the well is given by
$\left[v_{M}\right] +1$.

In what follows we will refer to the Dunham's expansion for
vibrational energies truncated to the cubic term. It is
convenient to write the energies in the form:
\begin{equation}
E^{\prime}(v)=E(v)/\hbar\omega= \left( v+\frac{1}{2}\right) -x_{e}\left(
v+\frac{1}{2}\right) ^{2}+y_{e}\left( v+\frac{1}{2}\right) ^{3}
\label{dunham3}
\end{equation}
where $\omega=2\pi c\omega_{e}$. The truncation is natural as the
experimental data show that each of the coefficients in the
expansion is much smaller than the preceding one, and furthermore
the contribution of terms beyond the cubic term is negligibly
small for all known diatomic molecules and seldom used in
practice \cite{{herz:dia},{OT}}.

The model we are proposing in the next section, aims at defining
a Hamiltonian which leads to Dunham-like formulae for the
molecular vibrations, using natural symmetry principles, starting
from an appropriate deformation of the Morse Hamiltonian.

\section{The Model and Its Approximations}

\subsection{Hamiltonian}

We shall use a Hamiltonian that can accommodate in different
approximations both the Morse energy (\ref{morse}) and the Dunham
expansion (\ref{dunham3}). This Hamiltonian is given by
\begin{equation}
\label{ham}H ~=~ \alpha\, (J_{+} J_{-} + J_{-} J_{+} )
\end{equation}
where $J_{+},J_{-}$ are the raising and lowering generators of
the quantum group deformation of the algebra $su(2)$, and
$\alpha$ is a constant which we shall fix below. In order to be
close to the usual formalism we shall realise these generators in
terms of anharmonic $q $-bosons. Their algebra, known as quantum
oscillator algebra\ $HW_{q}$\ or $q$-boson algebra, has been
introduced in
\cite{arik-coon:jmp,biedenharn:jphysa,macfarlane:jphysa}, and is
a generalisation of the Heisenberg-Weyl algebra obtained by
introducing a deformation parameter $q$. The algebra is defined
by:
\begin{equation}
aa^{\dagger} - q^{-1}a^{\dagger}a =q^{\hat{n}}\ ,\quad\lbrack\hat
{n},a]=-a\ ,\quad\lbrack\hat{n},a^{\dagger}]=a^{\dagger} \label{qanh}
\end{equation}
where $q$ is in general a complex number. This number is called
deformation parameter since the boson commutation relations of
the harmonic oscillator may be recovered for the value $q=1$. The
realization of $U_{q}(su(2))$ is taken from \cite{GP}:
\begin{equation}
\label{gp}J_{+} = a^{+} [2j-\hat{n}]_{q}\, , \quad J_{-} = a , \quad J_{0} =
\hat{n}-j
\end{equation}
where the $q$-number is defined as:
\[
[z]_{q}\, \equiv\frac{q^{z} -q^{-z}}{q-q^{-1}}\quad.
\]
In general $j$ may be an arbitrary complex number, but in our
context we shall take it to be real. These formulae indeed
produce the standard relations:
\begin{equation}
\label{sut}[J_{0} , J_{\pm}] = \pm J_{\pm}, \qquad[J_{+},J_{-}] = [2J_{0}
]_{q}\,
\end{equation}

In general the basis of our system is determined by the
application of the raising operator ~$a^{\dagger}$~ on the
vacuum. The latter is denoted by ~$|0\rangle$~ and is
characterised by the standard properties - anninihilation by the
lowering operator ~$a$~ and being eigenvector of the number
operator:
\begin{equation}
\label{vac}a\,|0\rangle\,=0\ ,\qquad\hat{n}\,|0\rangle\,=\, \nu\, |0\rangle\,
,
\end{equation}
where $\nu$ may be an arbitrary complex number. The basis
explicitly is:
\begin{equation}
\label{bas}|n\rangle\ \equiv\ (a^{\dagger})^{n}\,|0\rangle\,.
\end{equation}
The action of the quantum group $U_{q}(su(2))$ on this basis is:
\begin{align}
& J_{0}\,|n\rangle\ =\ (n+\nu-j)\, |n\rangle\,\nonumber\label{action}\\
& J_{-}\,|n\rangle\ =\ q^{\nu}\,[n]_{q}\,|n-1\rangle\,\nonumber\\
& J_{+}\,|n\rangle\ =\ [2j-\nu-n]_{q}\, |n+1\rangle.
\end{align}
For $j$ a non-negative (half-)integer and $\nu=0$ formulae
(\ref{gp}) realise a unitary irreducible representation of
$U_{q}(su(2))$ of dimension $2j+1$. To be close to this case
below we shall suppose ~$\nu=0$.

Substituting (\ref{gp}) in (\ref{ham}) we obtain:
\begin{equation}
\label{hama}H \ =\ \alpha\ \Big(
[2j]_{q}\, ( [2]_{q}\, [\hat{n}]_{q}\, q^{\epsilon
\hat{n}} +1) - [2]_{q}\, [\hat{n}]_{q}^{2}\, q^{\epsilon2j}\ \Big)
\end{equation}
where $\epsilon=\pm1$ and we have used
\begin{equation}
\label{add}a a^{+} = [\hat{n}+1]_{q}\,, \qquad a^{+} a = [\hat{n}]_{q}\,
\end{equation}
and the $q$-summation formula:
\begin{equation}
\label{qsum}[A+B]_{q}\, = [A]_{q}\, q^{\epsilon B} + [B]_{q}\, q^{-\epsilon
A}\ .
\end{equation}
Using again (\ref{qsum}) we rewrite (\ref{hama}) to obtain:
\begin{equation}
\label{hamba} H
\ =\ \alpha\ \Big( [\hn]_{q}\,[2j-\hn]_{q}\,[2]_{q} + [2j]_{q}
\Big)
\end{equation}
The action of this Hamiltonian on our basis is:
\begin{equation}
\label{acth}H\, |n\rangle\ =\ \mathcal{E}(n)\, |n\rangle\ =\ \alpha\,
\Big(
[n]_{q}\,[2j-n]_{q}\,[2]_{q} + [2j]_{q} \Big) \, |n\rangle\ .
\end{equation}

One motivation for (\ref{ham}) is that for $q=1$ and choosing
$\alpha= \hbar\omega/ 4j$ we get from (\ref{acth}) (essentially)
the Morse case:
\begin{equation}
\label{mors}\mathcal{E}(n)_{q=1} = \hbar\omega(n +{\textstyle {\frac{1}{2}}}
- n^{2}/2j)
\end{equation}
Note that this expression has a local maximum at $n=j$ which is:
\begin{equation}
\label{maxx}\left( \mathcal{E}_{q=1}\right) _{\mathrm{max}} \ =\ \mathcal{E}
(j)_{q=1}\ =\ \hbar\omega\, \frac{j+1}{2}
\end{equation}
and the bound levels are below this value of $j$, i.e., we have the
restriction:
\begin{equation}
\label{restr}n = 0,1, \dots, [j]
\end{equation}
Comparing this with the Morse case we identify $j$ with $v_{M}$ in
(\ref{mlevel}) which leads to the relation:
\begin{equation}
\label{reljx}\frac{1}{x_{e}} = 2j+1
\end{equation}

This limiting case prompts us to choose in the general case: $\alpha=
\hbar\omega/ [2]_{q}\, [2j]_{q}\,$ and then we have:
\begin{equation}
\label{hamb}H^{\prime}= H/(\hbar\omega) =
\frac{[\hn]_q\,[2j-\hn]_q}{[2j]_q} + \frac{1}{[2]_q}
\end{equation}
Then the eigenvalues are:
\eqn{eigen}
\ce'(n) \ =\ \ce(n)/(\hbar \omega) \ =\
\frac{[n]_q\,[2j-n]_q}{[2j]_q} + \frac{1}{[2]_q}
\ =\ \frac{ \sinh (n/p)
\sinh ((2j-n)/p)}{ \sinh (1/p) \sinh (2j/p)}  + \frac{1}{2\cosh
(1/p)}
\end{equation}
where we have introduced a parameter $p$ so that  $q = e^{-1/p}$.

This eigenvalue function is similar to the one given in formula
(25) of \cite{BRF}, though the approach in \cite{BRF} is not the
same to the present one, (a pair of $q$-bosons is used, the
Hamiltonian is different), and furthermore, we note that in
\cite{BRF} formula (25) does not follow (for $q\neq 1$) from the
Hamiltonian given in formula (24).

Note that in this form it is transparent that the eigenvalue
function has only one extremum as the extrema condition is:
\eqn{xetr}
\frac{\sinh((2j-2n)/p)}{p \sinh (1/p) \sinh (2j/p)} = 0
\end{equation}
and this is a local maximum at $ n=j$ (as in the
undeformed Morse case (\ref{mors})), the maximal value being:
\eqn{maxq} \ce_{\rm max} = \ce'(j) =
\left( [j]_q + 1\right) \frac{1}{2\cosh
(1/p)}
\end{equation}
which is, of course, a deformation of (\ref{maxx}).

\subsection{The case of positive molecular constant $y_{e}$}

One way to use this Hamiltonian is to suppose that $q$ is real and
close to 1. Then we expand to second order in $1/p$~: ~$\ q =
e^{-1/p} = 1 -1/p + 1/2p^{2} + \cdots$ and:
\begin{equation}
\label{appr}q^{m} = 1 -m/p + m^{2}/2p^{2} + \dots\ ,
\qquad[m]_{q}\, = \frac{\sinh(m/p)}{\sinh(1/p)} = m( 1 +
(m^{2}-1)/6p^{2}) + \cdots
\end{equation}
Further, one would consider the eigenvalues of $\hn$ which are
significantly smaller than $p$ and apply (\ref{appr}) for $m\to
\hn$. However, using this directly would not give exactly the
Dunham expansion. In order to be closer to the Dunham expansion we
expand the eigenvalue function (\ref{eigen}) as a function of $n $
around $-1/2$ (as in \cite{BRF}) and obtain:
\begin{equation}
\label{hambac}\mathcal{E}^{\prime}(n) = (1 - \sinh((2j+1/2)/p)) / \sinh(2j/p)
) /2 \cosh(1/2p) +\Big( (n + 1/2) \sinh((2j+1)/p) / p -
\end{equation}
\[
- (n + 1/2)^{2} \cosh((2j+1)/p) / p^{2} + 2 (n + 1/2)^{3} \sinh((2j+1)/p) / 3
p^{3} - \cdots\Big)  / \sinh(1/p) \sinh(2j/p)
\]

This expansion is also appropriate in the region when $n +1/2$ is
much smaller than $p$. Note that we make no restriction on the
values of $j$, an assumption which turns out to be justified. We
also see that the coefficient of $(n + 1/2)^{3}$ is positive which
means that this expression corresponds to the case when the
molecular constant (Dunham coefficient) $y_{e}$ is positive. The
cases of negative molecular constant $y_{e}$ will be considered
below.

We consider the expression in (\ref{hambac}) truncated to a cubic
polynomial in $(n + 1/2)$. The quadratic equation for its extrema
has the following solutions:
\begin{equation}
\label{extr}n_{\pm}= \frac{p}{2} \left( \coth((2j+1)/p) \pm\sqrt{\left(
\coth((2j+1)/p)) \right) ^{2} -2 }\, \right) - 1/2
\end{equation}
One is inclined to require that the discriminant be strictly
positive:
\[
\left( \tanh\frac{2j+1}{p}\right) ^{2} < \frac{1}{2}
\]
or
\begin{equation}
\label{newr}\frac{2j+1}{p} ~<~ \frac{1}{2} \ln(3+2\sqrt{2}) ~\approx~ 0.88
\end{equation}
In that case there are two extrema. The extremum at $n_{-}$ is
a local maximum. Obviously the value $n_{-}$ is a deformation of
the Morse value $j$, and if ~$j << p$~ one can expand:
\begin{equation}
\label{jdef}n_{-} = j + \frac{(2j+1)^{3}}{12p^{2}} + \cdots
\end{equation}
The extremum at $n_{+}$ is a local minimum. The value $n_{+} $
grows rapidly with $p$:
\[
n_{+} = \frac{p^{2}}{2j+1} -1/2 + \mathrm{const.}\frac{1}{p^{2}} + \cdots
\]
Thus, the eigenvalues $n$ cannot be near $n_{+} $, and one cannot
use the potential well around it.

Thus, when the restriction (\ref{newr}) holds, and similar to the
Morse case, we shall be interested in the region:
\begin{equation}
\label{resta}n = 0,1, \dots, [n_{-}]
\end{equation}

However, $j$ and $p$ will be determined by the experimental data
and we shall see that there are cases when (\ref{newr}) does not
hold. In that case, the potential does not have a local maximum,
i.e., there are cases when the Dunham expansion has a different
behaviour than the Morse expansion. Yet we are interested in
small values of $n$ when the two potentials do not differ
significantly. Thus, in these cases we shall be interested in the
Morse region (\ref{restr}): $n = 0,1, \dots, [j]$, though $j$
shall be determined from the Dunham expansion.

We would like now to check how the expansion (\ref{hambac})
corresponds to the Dunham expansion, which can be done
independently of whether (\ref{newr}) holds or not.

First, we take the ratio of the coefficients of the linear to
cubic terms in (n+1/2) both in the Dunham expansion
(\ref{dunham3}) and in our (\ref{hambac} ), and we find:
\begin{equation}
\label{lann}\frac{1}{y_{e}} = \frac{3}{2} p^{2}
\end{equation}
This means that $p$ is determined from the value of $y_{e}\,$:
\begin{equation}
\label{lannn}p = \sqrt{\frac{2}{3y_{e}}}
\end{equation}
Next we take the ratio of the coefficients of the linear to
quadratic terms in (n+1/2) both in the Dunham expansion
(\ref{dunham3}) and in our (\ref{hambac} ), and now we get:
\begin{equation}
\label{lapp}\frac{1}{x_{e}} = p\, \tanh\frac{2j+1}{p}
\end{equation}
This is a deformation of the usual relation of the Morse model
(\ref{reljx}), and if ~$j << p$~ one can expand:
\[
\frac{1}{x_{e}} = 2j+1 - \frac{(2j+1)^{3}}{3p^{2}} + ...
\]
Naturally, when $p\to\infty$ ($y_e\to0$), the Morse case applies.

More important is that (\ref{lapp}) gives a test for the
applicability of our model. Indeed, recall that the function
$\tanh$ is restricted: ~$\tanh z ~<~ 1$ for any real $z$. Thus,
we have:
\[
1 > \tanh\frac{2j+1}{p} = \frac{1}{p\,x_{e}} = \sqrt{\frac{3y_{e}}{2x_{e}^{2}}
}
\]
Thus, we have the following restriction on our model from the
experimental values $x_{e}$ and $y_{e}$:
\begin{equation}
\label{restrr}\frac{y_{e}}{x_{e}^{2}} < \frac{2}{3}
\end{equation}
Note that the limit ~$\frac{3y_{e}}{2x_{e}^{2}} \to1$~ with
values below 1 corresponds to the limit ~$j \to\infty$. Thus the
breaking point ~$\frac{y_{e}}{x_{e}^{2}} = \frac{2}{3}$~ means
$j=\infty$.

If the bound goes the other way around, i.e.,
~$\frac{y_{e}}{x_{e}^{2}} > \frac{2}{3}$, then our model is not
applicable (even if we consider complex $j$).

In the cases when (\ref{restrr}) holds from (\ref{lapp}) we
determine the value of $j$ using the value of $p$ from
(\ref{lannn}), i.e.,
\begin{equation}
\label{lappp}j = {\textstyle {\frac{1}{2}}} \left( \sqrt{\frac{2}{3y_{e}}
}\ \mathrm{arctanh}\left( \sqrt{\frac{3y_{e}}{2x_{e}^{2}}}\ \right) -1
\right)
\end{equation}

It is useful to write down the formula for ~$n_{-}$~ in terms of ~$x_{e}
,y_{e}$~ using (\ref{lannn}) and (\ref{lapp}):
\begin{equation}
\label{extry}n_{-} = \frac{x_{e}}{3y_{e}} \left( 1- \sqrt{ 1 - \frac{3y_{e}
}{x_{e}^{2}}}\ \right) - 1/2
\end{equation}
from which it is clear that the restriction (\ref{newr}) translates into:
\begin{equation}
\label{restry}\frac{y_{e}}{x_{e}^{2}} ~<~ \frac{1}{3}
\end{equation}
which is indeed stronger than (\ref{restrr}), but is not a
restriction on the applicability of our model.

In the case when $y_{e}/x_{e}^{2} << 1$ we can use the expansion:
\[
\mathrm{arctanh}\ z = z + z^{3} + \cdots
\]
and obtain from (\ref{lappp}):
\begin{equation}
\label{lapa}j = \frac{1}{2x_{e}} - \frac{1}{2} + \frac{3y_{e}}{4x_{e}^{3}} +
\cdots
\end{equation}

\subsection{The case of negative molecular constant $y_{e}$}

In order to accommodate the situation when the molecular constant
$y_{e}$ is negative we have to consider the deformation parameter
$q$ to be a phase (though not a root of unity), or equivalently
to make the replacement: $p\ \mapsto\ ip$, then $q\ \mapsto\ q =
e^{i/p}$. Almost everything may be obtained from what we have in
the case of real $q$ by this replacement. In particular, one
needs:
\[
\sinh z/p \mapsto-i \sin/p\ ,\quad[z]_{q} \mapsto\frac{\sin(z/p)}{\sin(1/p)}
\ ,\quad\cosh z/p \mapsto\cos z/p\ ,
\]
\[
(1/p) \sinh a/p \mapsto- (1/p) \sin a/p , \quad(1/p^{3}) \sinh a/p \mapsto+
(1/p^{3}) \sin a/p ,
\]
\[
\sinh(1/p) \sinh(a/p) \mapsto- \sin(1/p) \sin(a/p)
\]

We start with the analogues of (\ref{hamba}):
\begin{equation}
\label{hambi}H^{\prime}_{-} = [\hat{n}]_{q}\, [2j-\hat{n}]_{q}\, / [2j]_{q}\,
+ 1/[2]_{q}\ ,
\end{equation}
and (\ref{eigen})
\begin{equation}
\label{eigenm}\mathcal{E}^{\prime}_{-}(n) \ =\ = \frac{ \sin(n/p)
\sin((2j-n)/p)}{ \sin(1/p) \sin(2j/p)} + \frac{1}{2\cos(1/p)}
\end{equation}
This function is very different from (\ref{eigen}), in
particular, it has an infinite number of extrema since the
extremum condition is:
\begin{equation}
\label{xeti}\frac{\sin((2j-2n)/p)}{p \sin(1/p) \sin(2j/p)} = 0
\end{equation}
i.e., the extrema are at: $n = j + kp\pi/2 $,
$k=0,\pm1,\pm2,...$. However, again we would be interested in the
region restricted by the first positive local maximum:
\[
n = 0,1, \dots, [j]
\]
This is also consistent with our aim of fitting the Dunham
expansion. The maximal value is:
\begin{equation}
\label{maxi}\mathcal{E}^{\prime}_{-,\mathrm{max}} = \left( \frac{\sin
(j/p)}{\sin(1/p)} + 1\right) \frac{1}{2\cos(1/p)}
\end{equation}
which is a deformation of (\ref{maxx}).

The expansion of $n$ around $-1/2$ is:
\begin{equation}
\label{hambaci}\mathcal{E}^{\prime}_{-}(n) = (1 + \sin((2j+1/2)/p) /
\sin(2j/p) ) /2 \cos(1/2p) +\Big( (n + 1/2) \sin((2j+1)/p) / p -
\end{equation}
\[
- (n + 1/2)^{2} \cos((2j+1)/p) / p^{2} - 2 (n + 1/2)^{3} \sin((2j+1)/p) / 3
p^{3} - \cdots\Big)  / \sin(1/p) \sin(2j/p)
\]
i.e., this is suitable for the Dunham expansion with negative
$y_{e}$. We consider the expansion in (\ref{hambaci}) truncated
to a cubic polynomial in $(n + 1/2)$. As such it has two extrema
at the points:
\begin{equation}
\label{extri}n_{\pm}= \frac{p}{2} \left( \mp\sqrt{\left( \cot((2j+1)/p)
\right) ^{2} +2 } ~-~ \cot((2j+1)/p) \right) - {\textstyle {\frac{1}{2}}}
\end{equation}
The value $n_{+}$ is negative and thus inaccessible. The extremum
at $n_{-} $ is a local maximum.

Further, we take the ratio of the coefficients of the
linear/cubic terms in (n+1/2) both in Dunham (\ref{dunham3}) and
in our (\ref{hambaci}) and we get:
\begin{equation}
\label{lani}\frac{1}{y_{e}} = -\frac{3}{2} p^{2} < 0
\end{equation}
This means that $p$ is determined from the value of $y_{e}\,$:
\begin{equation}
\label{lanni}p = \sqrt{-\frac{2}{3y_{e}}} \ , \quad y_{e} < 0
\end{equation}
Next we take the ratio of the coefficients of the
linear/quadratic terms in (n+1/2) both in Dunham (\ref{dunham3})
and in our (\ref{hambaci}) and we get:
\begin{equation}
\label{lapi}\frac{1}{x_{e}} = p\, \tan\frac{2j+1}{p}
\end{equation}
{}From this we should determine the value of $j$ using the value
of $p$ from (\ref{lanni}), however, further we have to
distinguish whether ~$x_{e}>0$~ (which holds in most cases) or
~$x_{e}<0$.

\vskip  5mm

$\bullet$ ~~$x_{e}>0$ ~$\Rightarrow~\tan\frac{2j+1}{p} ~>~ 0$

\vskip  3mm

Then we have for the value $n_{-}$~:
\begin{equation}
\label{extryi}n_{-} = \frac{x_{e}}{3|y_{e}|} \left( \, \sqrt{ 1 +
\frac{3|y_{e}|}{x_{e}^{2}}} -1 \right) - {\textstyle
{\frac{1}{2}}} \ , \qquad y_{e}<0 \ , ~~x_{e}>0 \ .
\end{equation}
It is a deformation of the Morse value, and if $j<<p$ we can expand:
\begin{equation}
\label{jdefi}n_{-} = j - \frac{(2j+1)^{3}}{12p^{2}} + \cdots
\end{equation}
Thus, similarly to the Morse case we shall be interested in the region:
\begin{equation}
\label{resti}n = 0,1, \dots, [n_{-}]
\end{equation}

The value of ~$j$~ is obtained from (\ref{lapi})~:
\begin{equation}
\label{lappi}j = {\textstyle {\frac{1}{2}}} \left( \sqrt{-\frac{2}{3y_{e}}
}\ \mathrm{arctan}\left( \sqrt{-\frac{3y_{e}}{2x_{e}^{2}}}\ \right) -1
\right) \ , \quad y_{e} < 0 \ , ~x_{e} > 0 \ .
\end{equation}
Since the function \ arctan\ is multivalued, in the last formula
we take the value which is closest to the Morse value from
(\ref{reljx}). Analogously, we use this for the following
expansion which is valid when ~$-y_{e} << x_{e} ^{2}$~:
\begin{equation}
\label{lapai}j = \frac{1}{2x_{e}} - \frac{1}{2} - \frac{3y_{e}}{4x_{e}^{3}} +
\cdots
\end{equation}

\vskip  5mm

$\bullet$ ~~$x_{e}<0$ ~$\Rightarrow~\tan\frac{2j+1}{p} ~<~ 0$

\vskip  3mm

In this case the relevant parameter is ~$j^{\prime}$~ which is
complementary to ~$j$~ w.r.t. ~$p\pi/2$. Then,
\[
\tan(2j+1)/p = \tan\left( \pi- (2j^{\prime}+1)/p \right) = - \tan
(2j^{\prime}+1)/p
\]
and instead of (\ref{lapi}) we shall use:
\begin{equation}
\label{lapiz}\frac{1}{|x_{e}|} = p\, \tan\frac{2j^{\prime}+1}{p}
\end{equation}
Further, for the value at which there is maximum we have:
\eqnn{extriz}
n_- &=& \frac{p}{2} \left( \sqrt{ \left(\cot
((2j'+1)/p) \right)^2 +2 } ~+~ \cot ((2j'+1)/p) \right) - \half ~=\cr
&=& \frac{|x_e|}{3|y_e|} \left( \sqrt{1 + \frac{3|y_e|}{x^2_e}
} +1 \right) - \half
\eea
It is not a deformation of the Morse value, since it grows with
$p$ and will not be useful for our purposes. Thus, similar to a
case above we shall be interested in the region:
\begin{equation}
\label{restiz}n = 0,1, \dots, [j^{\prime}]
\end{equation}
The value of ~$j^{\prime}$~ is obtained from (\ref{lapiz})~:
\begin{equation}
\label{lappiz}j^{\prime}= {\textstyle {\frac{1}{2}}} \left( \sqrt
{-\frac{2}{3y_{e}}}\ \mathrm{arctan}\left( \sqrt{-\frac{3y_{e}}{2x_{e}^{2}}
}\ \right) -1 \right) \ , \quad y_{e} < 0 \ , ~x_{e} < 0 \ .
\end{equation}
In the last formula we take the value which is closest to the
Morse value w.r.t. $j^{\prime}$, i.e., $j^{\prime}=
\frac{1}{2|x_{e}|} - \frac{1}{2}$. Analogously, we use this for
the following expansion which is valid when ~$-y_{e} <<
x_{e}^{2}$~:
\begin{equation}
\label{lapaiz}j^{\prime}= \frac{1}{2|x_{e}|} - \frac{1}{2} - \frac{3y_{e}
}{4x_{e}^{3}} + \cdots
\end{equation}

\section{Analysis of Experimental Data}

In the Tables below, we have calculated the independent
parameters of the model $p$ and $j$, using the values of the
molecular constants published in \cite{herz:dia} for 161
electronic states of diatomic molecules. The molecules are listed
in alphabetical order in the first column of the tables, the
corresponding electronic states are given in the second column.
For convenience, in the third and fourth column we display the
values of the anharmonic molecular constants
$x_{e}=\omega_{e}x_{e}/\omega_{e}$ and
$y_{e}=\omega_{e}y_{e}/\omega_{e}$ calculated from the published
data in \cite{herz:dia}. The values of the parameters of the
model $p$ and $j$ are given in the following two columns. In the
last column, where appropriate, we give the value of
$\left[n_{-}\right]$.

In Table 1 we display the cases when the molecular constant
$y_{e}$ is positive and the values of the parameters $p$ and $j$
are calculated from (\ref{lannn}) and (\ref{lappp}) respectively.
In the cases when $\ $ $y_{e}/\left( x_{e}\right) ^{2}<1/3,$ the
potential curve for a given electronic state has a local maximum
and the maximal number of vibrational levels is determined by the
value of $\ \left[ n_{-}\right]$ calculated from (\ref{extry}).

When $1/3<$ $y_{e}/\left( x_{e}\right) ^{2}<2/3,$ Dunham's
potential curve truncated to the cubic term does not have a local
extremum, which is indicated by ''x'' in the column for $\left[
n_{-}\right] $. In these cases the maximal number of vibrational
levels in the model is determined by the value $\left[ j\right]$.

\begin{table}
\caption{Parameters of the model for positive molecular constant
$y_e$}
\begin{tabular}
[c]{|c|c|c|c|c|c|c|}\hline molecule & state & $x_{e}$ & $y_{e}$ &
$p$ & $j$ & $\left[  n_{-}\right] $\\\hline $^{27}$Al$^{1}$H &
$X\;^{1}\Sigma^{+}$ & $\allowbreak1.\,\allowbreak
732\,17\times10^{-2}$ &
$\allowbreak1.\,\allowbreak545\,26\times10^{-4}$ &
$\allowbreak65.\,\allowbreak68$ & $\allowbreak44.\,\allowbreak53$
& x\\\hline $^{27}$Al$^{2}$H & $X\;^{1}\Sigma^{+}$ &
$\allowbreak1.\,\allowbreak 249\,06\times10^{-2}$ &
$\allowbreak8.\,\allowbreak086\,14\times10^{-5}$ &
$\allowbreak90.\,\allowbreak80$ & $\allowbreak62.\,\allowbreak31$
& x\\\hline
$^{9}$Be$^{16}$O & $%
\begin{array}
[c]{c}%
A\;^{1}\Pi\\
X\;^{1}\Sigma^{+}%
\end{array}
$ & $%
\begin{array}
[c]{c}%
\allowbreak7.\,\allowbreak353\,8\times10^{-3}\\
7.\,\allowbreak953\,69\times10^{-3}\allowbreak
\end{array}
$ & $%
\begin{array}
[c]{c}%
2.\,\allowbreak961\,80\times10^{-5}\\
1.\,\allowbreak502\,70\times10^{-5}%
\end{array}
$ & $%
\begin{array}
[c]{c}%
150.\,\allowbreak03\\
210.63
\end{array}
$ & $%
\begin{array}
[c]{c}%
112.\,\allowbreak54\\
71.\,\allowbreak99
\end{array}
$ &
\begin{tabular}
[c]{c}%
x\\
$81$%
\end{tabular}
\\\hline
$^{209}$Bi$^{35}$Cl & $B$ & $\allowbreak\allowbreak9.\,\allowbreak
338\,29\times10^{-3}$ &
$\allowbreak3.\,\allowbreak965\,3\times10^{-6}$ &
$\allowbreak410.\,\allowbreak03$ & $\allowbreak54.\,\allowbreak31$
& $55$\\\hline $^{11}$B$^{16}$O & $A\;^{2}\Pi_{i}$ &
$8.\,\allowbreak849\,85\times10^{-3}$ &
$\allowbreak3.\,\allowbreak886\,73\times10^{-5}$ & $\allowbreak
130.\,\allowbreak97$ & $\allowbreak84.90$ & x\\\hline Ca$^{35}$Cl
& $A\;^{2}\Pi$ & $\allowbreak3.\,\allowbreak178\,95\times10^{-3}$
& $\allowbreak3.\,\allowbreak288\,57\times10^{-6}$ & $\allowbreak
450.\,\allowbreak25$ & $\allowbreak194.\,\allowbreak16$ &
$\allowbreak 272$\\\hline $^{12}$C$^{16}$O & $X\;^{1}\Sigma^{+}$ &
$\allowbreak6.\,\allowbreak 177\,01\times10^{-3}$ &
$\allowbreak1.\,\allowbreak413\,36\times10^{-5}$ &
$\allowbreak217.\,\allowbreak18$ & $104.\,\allowbreak02$ &
x\\\hline ($^{12}$C$^{16}$O)$^{+}$ & $A\;^{2}\Pi_{i}$ &
$\allowbreak8.\,\allowbreak 655\,24\times10^{-3}$ &
$\allowbreak8.\,\allowbreak386\,36\times10^{-6}$ & $281.95$ &
$\allowbreak60.\,\allowbreak87$ & $63$\\\hline $^{1}$H$_{2}$ &
\begin{tabular}
[c]{c}%
$d$\ $^{3}\Pi_{u}$\\
$a\;^{3}\Sigma_{u}^{\;+}$\\
$X\;^{1}\Sigma_{g}^{\;+}$%
\end{tabular}
& $%
\begin{array}
[c]{c}%
\allowbreak2.\,\allowbreak794\,34\times10^{-2}\\
\allowbreak2.\,\allowbreak688\,73\times10^{-2}\\
\allowbreak\allowbreak2.\,\allowbreak684\,52\times10^{-2}%
\end{array}
$ & $%
\begin{array}
[c]{c}%
\allowbreak3.\,\allowbreak710\,61\times10^{-4}\\
\allowbreak3.\,\allowbreak452\,38\times10^{-4}\\
\allowbreak6.\,\allowbreak598\,11\times10^{-5}%
\end{array}
$ & $%
\begin{array}
[c]{c}%
\allowbreak42.\,\allowbreak39\\
\allowbreak43.\,\allowbreak94\\
\allowbreak100.\,\allowbreak52
\end{array}
$ & $%
\begin{array}
[c]{c}%
\allowbreak25.\,\allowbreak69\\
\allowbreak26.\,\allowbreak82\\
19.\,\allowbreak06
\end{array}
\allowbreak$ &
\begin{tabular}
[c]{c}%
x\\
x\\
$19$%
\end{tabular}
\\\hline
$^{1}$H$^{2}$H &
\begin{tabular}
[c]{c}%
$d$\ $^{3}\Pi_{u}$\\
$e\;^{3}\Sigma_{u}^{\;+}$\\
$a\;^{3}\Sigma_{g}^{\;+}$\\
$X\;^{1}\Sigma_{g}^{+}$%
\end{tabular}
&
\begin{tabular}
[c]{c}%
$\allowbreak2.\,\allowbreak420\,92\times10^{-2}$\\
$\allowbreak2.\,\allowbreak713\,67\times10^{-2}$\\
$2.\,\allowbreak329\,28\times10^{-2}$\\
$\allowbreak2.\,\allowbreak487\,71\times10^{-2}$%
\end{tabular}
&
\begin{tabular}
[c]{c}%
$2.\,\allowbreak822\,95\times10^{-4}$\\
$\allowbreak2.\,\allowbreak739\,91\times10^{-4}$\\
$2.\,\allowbreak599\,16\times10^{-4}$\\
$\allowbreak\allowbreak3.\,\allowbreak816\,78\times10^{-4}$%
\end{tabular}
&
\begin{tabular}
[c]{c}%
$\allowbreak48.\,\allowbreak60$\\
$\allowbreak49.\,\allowbreak33$\\
$\allowbreak50.\,\allowbreak64$\\
$41.\,\allowbreak79$%
\end{tabular}
$\,\allowbreak\allowbreak$ & $%
\begin{array}
[c]{c}%
30.\,\allowbreak02\\
\allowbreak23.\,\allowbreak33\\
\allowbreak31.\,\allowbreak10\\
\allowbreak40.\,\allowbreak66
\end{array}
\allowbreak$ & $%
\begin{array}
[c]{c}%
\text{x}\\
\text{x}\\
\text{x}\\
\text{x}%
\end{array}
$\\\hline $^{2}$H$_{2}$ &
\begin{tabular}
[c]{c}%
$d$\ $^{3}\Pi_{u}$\\
$e\;^{3}\Sigma_{u}^{\;+}$\\
$a\;^{3}\Sigma_{g}^{\;+}$\\
$D\;^{1}\Pi_{u}$%
\end{tabular}
& $%
\begin{array}
[c]{c}%
\allowbreak1.\,\allowbreak962\,79\times10^{-2}\\
2.\,\allowbreak216\,95\times10^{-2}\\
\allowbreak1.\,\allowbreak906\,84\times10^{-2}\\
\allowbreak1.\,\allowbreak855\,36\times10^{-2}%
\end{array}
$ & $%
\begin{array}
[c]{c}%
1.\,\allowbreak430\,09\times10^{-4}\\
1.\,\allowbreak843\,71\times10^{-4}\\
\allowbreak1.\,\allowbreak802\,91\times10^{-4}\\
\allowbreak3.\,\allowbreak606\,96\times10^{-5}%
\end{array}
$ & $\allowbreak\allowbreak%
\begin{array}
[c]{c}%
68.\,\allowbreak28\\
\allowbreak60.\,\allowbreak13\\
60.\,\allowbreak81\\
135.95
\end{array}
$ & $%
\begin{array}
[c]{c}%
32.\,\allowbreak42\\
\allowbreak28.\,\allowbreak76\\
39.\,\allowbreak11\\
\allowbreak28.\,\allowbreak01
\end{array}
$ & $%
\begin{array}
[c]{c}%
\text{x}\\
\text{x}\\
\text{x}\\
28
\end{array}
$\\\hline $^{1}$H$^{3}$H &
\begin{tabular}
[c]{c}%
$d\;^{3}\Pi_{u}$\\
$a\;^{3}\Sigma_{g}^{\;+}$%
\end{tabular}
&
\begin{tabular}
[c]{c}%
$\allowbreak2.\,\allowbreak242\,67\times10^{-2}$\\
$\allowbreak2.\,\allowbreak197\,51\times10^{-2}$%
\end{tabular}
&
\begin{tabular}
[c]{c}%
$\allowbreak2.\,\allowbreak369\,73\times10^{-4}$\\
$2.\,\allowbreak305\,91\times10^{-4}$%
\end{tabular}
&
\begin{tabular}
[c]{c}%
$53.\,\allowbreak04$\\
$\allowbreak53.\,\allowbreak77$%
\end{tabular}
&
\begin{tabular}
[c]{c}%
$31.\,\allowbreak95$\\
$32.\,\allowbreak92$%
\end{tabular}
&
\begin{tabular}
[c]{c}%
x\\
x
\end{tabular}
\\\hline
$^{3}$H$_{2}$ & $d\;^{3}\Pi_{u}$ &
$\allowbreak1.\,\allowbreak613\,21\times 10^{-2}$ $\allowbreak$ &
$\allowbreak1.\,\allowbreak158\,80\times 10^{-4}\allowbreak$ &
$\allowbreak75.\,\allowbreak85$ & $\allowbreak43.06$ & x\\\hline
$^{1}$H$^{35}$Cl & $X\;^{1}\Sigma^{+}$ &
$\allowbreak1.\,\allowbreak 740\,95\times10^{-2}$ &
$1.\,\allowbreak873\,07\times10^{-5}$ & $\allowbreak
188.\,\allowbreak66$ & $\allowbreak29.\,\allowbreak16$ &
$29$\\\hline $\left(  ^{1}\text{H}^{35}\text{Cl}\right)  ^{+}$ &
$A\;^{2}\Sigma^{+}$ &
$\allowbreak2.\,\allowbreak464\,83\times10^{-2}$ &
$\allowbreak2.\,\allowbreak 129\,79\times10^{-4}$ &
$\allowbreak55.\,\allowbreak95$ & $25.\,\allowbreak19$ & x\\\hline
$^{1}$H$^{19}$F & $X\;^{1}\Sigma^{+}$ &
$\allowbreak2.\,\allowbreak 176\,4\times10^{-2}$ &
$\allowbreak2.\,\allowbreak368\times10^{-4}$ &
$\allowbreak53.\,\allowbreak06$ & $\allowbreak34.\,\allowbreak43$
& x\\\hline Hg$^{2}$H & $X\;^{2}\Sigma$ &
$\allowbreak5.\,\allowbreak017\,33\times10^{-2}$ &
$\allowbreak1.\,\allowbreak118\,42\times10^{-3}$ & $\allowbreak
24.\,\allowbreak41$ & $\allowbreak13.\,\allowbreak49$ & x\\\hline
$^{127}$I$_{2}$ & $A\;^{3}\Pi_{u}$ & $\allowbreak2.\,\allowbreak
272\,73\times10^{-2}$ & $1.\,\allowbreak818\,18\times10^{-4}$ &
$\allowbreak 60.\,\allowbreak55$ & $\allowbreak27.\,\allowbreak40$
& x\\\hline $^{39}$K$_{2}$ & $D\;\left(  ^{1}\Pi_{u}\right)  $ &
$1.\,\allowbreak 461\,04\times10^{-2}$ &
$1.\,\allowbreak623\,38\times10^{-5}$ &
$202.\,\allowbreak65\allowbreak$ $\allowbreak$ &
$30.\,\allowbreak26$ $\allowbreak$ & $35$\\\hline KBr &
$X\;^{1}\Sigma^{+}$ &
$\allowbreak3.\,\allowbreak030\,3\times10^{-3}$ &
$\allowbreak4.\,\allowbreak761\,9\times10^{-6}$ & $\allowbreak
374.\,\allowbreak12$ & $174.\,\allowbreak08$ & x\\\hline KCl &
$X\;^{1}\Sigma^{+}$ &
$\allowbreak3.\,\allowbreak214\,29\times10^{-3}$ &
$\allowbreak3.\,\allowbreak928\,57\times10^{-6}$ & $\allowbreak
411.\,\allowbreak94$ & $164.\,\allowbreak61$ & x\\\hline
K$^{127}$I & $X\;^{1}\Sigma^{+}$ &
$\allowbreak3.\,\allowbreak301\,89\times 10^{-3}$ &
$4.\,\allowbreak716\,98\times10^{-6}$ & $375.\,\allowbreak94$ &
$\allowbreak208.\,\allowbreak96$ & x\\\hline $^{7}$Li$_{2}$ &
$A\;^{1}\Sigma_{u}^{+}$ & $6.\,\allowbreak161\,68\times 10^{-3}$ &
$\allowbreak7.\,\allowbreak046\,39\times10^{-6}$ & $\allowbreak
307.\,\allowbreak59$ & $89.\,\allowbreak76$ & $96$\\\hline
$^{7}$Li$^{1}$H & $X\;^{1}\Sigma^{+}$ &
$1.\,\allowbreak650\,48\times10^{-2}$ &
$1.\,\allowbreak161\,74\times10^{-4}$ &
$\allowbreak75.\,\allowbreak75$ & $41.\,\allowbreak09$ & x\\\hline
Li$^{127}$I & $X\;^{1}\Sigma^{+}$ &
$3.\,\allowbreak333\,33\times10^{-3}$ &
$3.\,\allowbreak777\,78\times10^{-6}$ &
$\allowbreak420.\,\allowbreak08$ & $187.\,\allowbreak61$ &
x\\\hline $^{14}$N$_{2}$ & $X\;^{1}\Sigma_{g}^{+}$ &
$6.\,\allowbreak126\,44\times 10^{-3}$ &
$\allowbreak3.\,\allowbreak182\,73\times10^{-6}$ &
$457.\,\allowbreak67$ & $\allowbreak84.\,\allowbreak86$ &
$\allowbreak 87$\\\hline $^{23}$NaBr & $X\;^{1}\Sigma^{+}$ &
$3.\,\allowbreak650\,79\times10^{-3}$ &
$\allowbreak2.\,\allowbreak539\,68\times10^{-6}$ & $\allowbreak
512.\,\allowbreak35$ & $\allowbreak152.\,\allowbreak33$ &
$165$\\\hline $^{23}$Na$^{1}$H & $X\;^{1}\Sigma^{+}$ &
$1.\,\allowbreak682\,31\times10^{-2}$ &
$\allowbreak1.\,\allowbreak364\,95\times10^{-4}$ & $\allowbreak
69.\,\allowbreak89$ & $\allowbreak43.\,\allowbreak46$ & x\\\hline
$^{23}$Na$^{127}$I & $X\;^{1}\Sigma^{+}$ &
$\allowbreak2.\,\allowbreak 622\,38\times10^{-3}$ &
$\allowbreak3.\,\allowbreak496\,5\times10^{-6}$ &
$\allowbreak436.\,\allowbreak65$ & $293.\,\allowbreak55$ &
x\\\hline $^{16}$O$_{2}$ & $X\;^{3}\Sigma_{g}^{-}$ &
$\allowbreak7.\,\allowbreak 639\,39\times10^{-3}$ &
$\allowbreak3.\,\allowbreak454\,91\times10^{-5}$ &
$\allowbreak138.\,\allowbreak91$ & $121.\,\allowbreak64$ &
x\\\hline Rb$^{1}$H & $X\;^{1}\Sigma^{+}$ &
$1.\,\allowbreak510\,51\times10^{-2}$ &
$\allowbreak8.\,\allowbreak006\,23\times10^{-5}$ &
$91.\,\allowbreak25$ & $\allowbreak41.\,\allowbreak44$ & x\\\hline
$^{28}$Si$^{14}$N & $B\;^{2}\Sigma^{+}$ &
$1.\,\allowbreak623\,86\times 10^{-2}$ &
$\allowbreak1.\,\allowbreak136\,95\times10^{-4}$ &
$76.\,\allowbreak57$ & $42.\,\allowbreak01$ & x\\\hline
$^{28}$Si$^{16}$O & $X\;^{1}\Sigma^{+}$ &
$\allowbreak4.\,\allowbreak 868\,64\times10^{-3}$ &
$\allowbreak2.\,\allowbreak648\,89\times10^{-6}$ $\,$ &
$\allowbreak\allowbreak501.\,\allowbreak68$ &
$\allowbreak108.\,\allowbreak 59$ & $\allowbreak112$\\\hline
Sr$^{19}$F & $A\;^{2}\Pi$ & $4.\,\allowbreak467\,3\times10^{-3}$
$\allowbreak$ &
$\allowbreak3.\,\allowbreak953\,4\times10^{-6}\allowbreak$
$\allowbreak$ & $\allowbreak410.\,\allowbreak65$ &
$\allowbreak125.\,\allowbreak03$ $\allowbreak$ & $136$\\\hline
Zn$^{1}$H & $X\;^{2}\Sigma^{+}$ & $\allowbreak0.0\,\allowbreak343$
& $2.\,\allowbreak475\,7\times10^{-4}$ &
$\allowbreak51.\,\allowbreak89$ & $\allowbreak15.\,\allowbreak99$
& $\allowbreak17$\\\hline
\end{tabular}
\end{table}

As discussed in the previous section, when $y_{e}/\left(
x_{e}\right)^{2}>2/3$, the model cannot be applied in its present
form. The experimental data in \cite{herz:dia} show that this
condition is satisfied for 30 electronic states of the following
diatomic molecules:
~$^{75}$As$_{2}$ state $X\ ^{1}\Sigma_{g}^{+}$, $^{11}$B$^{79}$Br state
$A\;^{1}\Pi$, $^{209}$Bi$^{19}$F state $A$, $^{12}$C$_{2}$ state $B\;^{3}%
\Pi_{g}$, $^{40}$Ca$^{12}$F\ state $A\;^{2}\Pi$,\ $^{1}$H$_{2}$ state
$B\;^{1}\Sigma_{u}^{+}$, $^{1}$H$^{2}$H states $C\;^{1}\Pi_{u}$ and
$B\ ^{1}\Sigma_{u}^{+}$, $^{2}$H$_{2}$ states $C\;^{1}\Pi_{u}$, $B\ ^{1}%
\Sigma_{u}^{+}$ and $X\;^{1}\Sigma_{g}^{+}$, $^{3}$H$_{2}$ state
$a\;^{3}\Sigma_{g}^{+}$, Hg$^{35}$Cl state $X\;^{2}\Sigma^{+}$, $\left(
\text{Hg}^{1}\text{H}\right)  ^{+}$ state $X\ ^{1}\Sigma^{+}$, $\left(
\text{Hg}^{2}\text{H}\right)  ^{+}$ state $X\ ^{1}\Sigma^{+}$,\ $^{113}%
$In$^{1}$H state $X\ ^{1}\Sigma^{+}$,\ $^{39}$K$_{2}$ state $B\;^{1}\Pi_{u}%
$,\ $^{7}$Li$^{2}$H state $X\ ^{1}\Sigma^{+}$, $^{(24)}$Mg$^{1}$H state
$A\;^{2}\Pi_{r}$, $^{55}$Mn$^{16}$O state $A$, $^{23}$NaK state $C\;^{1}\Pi$,
$^{14}$N$^{16}$O state $B\;^{2}\Pi_{r}$,\ $^{31}$P$_{2}$ state $A\;^{1}%
\Sigma_{u}^{+}$,\ Pb$_{2}$ state\ $A$,\ $^{80}$Se$_{2}$ state\ $\left(
X\right)  \;\left(  ^{1}\Sigma^{+}\right)  $,\ $^{28}$Si$^{16}$O$\;$state
$A\;^{1}\Pi$, $^{28}$SiTe state $E$, YbCl state $B\;\left(  ^{2}\Pi\right)  $
and $^{90}$Zr$^{16}$O state $a\;\left(  ^{1}\Sigma\right)  $.

In Table 2 we show the cases when the molecular constant $y_{e}$
is negative and the values of the parameters $p$ and $j$ are
calculated from (\ref{lanni}) and (\ref{lappi}) respectively.
Here, the model works without any restrictions. The number of
vibrational levels is determined by $\left[ n_{-}\right] \;$
calculated from (\ref{extryi}). The experimental data shows that
when $n\_\sim j$, the value of $j$ is at least three times
smaller than the value of $p$.

\begin{table}
\caption{Parameters of the model for negative molecular constant
$y_e$}
\begin{tabular}
[c]{|c|c|c|c|c|c|c|}\hline molecule & state & $x_{e}$ & $\left|
y_{e}\right| $ & $p$ & $j$ & $\left[ n_{-}\right]  $\\\hline
$^{109}$Ag$^{81}$Br & $B\;\left(  ^{3}\Pi_{0}^{+}\right) $ &
$2.\,\allowbreak 461\,3\times10^{-2}$ &
$\allowbreak3.\,\allowbreak318\,6\times10^{-4}$ &
$\allowbreak44.\,82$ & $\allowbreak16.\,\allowbreak00$ &
$14$\\\hline $^{109}$Ag$^{35}$Cl & $B\;\left(
^{3}\Pi_{0}^{+}\right) $ & $\allowbreak
2.\,\allowbreak135\,2\times10^{-2}$ &
$3.\,\allowbreak380\,8\times10^{-4}$ &
$\allowbreak44.\,\allowbreak41$ & $\allowbreak17.\,\allowbreak53$
& $16$\\\hline $^{109}$Ag$^{127}$I & $B\;\left(
^{3}\Pi_{0}^{+}\right) $ & $1.\,\allowbreak
286\,6\times10^{-2}\allowbreak$ &
$\allowbreak3.\,\allowbreak908\,8\times 10^{-3}$ &
$\allowbreak13.\,\allowbreak06$ & $\allowbreak8.\,\allowbreak67$ &
$\allowbreak7$\\\hline $^{27}$Al$^{79}$Br & $A\;^{1}\Pi$\  &
$\allowbreak2.\,\allowbreak 153\,4\times10^{-2}$ &
$\allowbreak1.\,\allowbreak773\,2\times10^{-3}$ &
$\allowbreak19.\,\allowbreak39$ & $\allowbreak10.\,\allowbreak90$
& $\allowbreak9$\\\hline $^{27}$Al$^{35}$Cl & $A\;^{1}\Pi$ &
$9.\,\allowbreak712\times10^{-3}$ &
$4.\,\allowbreak800\,4\times10^{-4}$ & $37.\,\allowbreak27$ &
$\allowbreak 22.\,\allowbreak30$ & $19$\\\hline $^{27}$Al$^{19}$F
& $A\;^{1}\Pi$ & $1.\,\allowbreak032\,9\times10^{-2}$ &
$\allowbreak2.\,\allowbreak272\,5\times10^{-4}$ &
$54.\,\allowbreak16$ & $\allowbreak28.\,\allowbreak23$ &
$25$\\\hline $^{197}$Au$^{1}$H &
\begin{tabular}
[c]{c}%
$A\;^{1}\Sigma^{+}$\\
$X\;^{1}\Sigma^{+}$%
\end{tabular}
&
\begin{tabular}
[c]{c}%
$\allowbreak3.\,\allowbreak297\,89\times10^{-2}$\\
$1.\,\allowbreak870\,7\times10^{-2}$%
\end{tabular}
&
\begin{tabular}
[c]{c}%
$2.\,\allowbreak353\,93\times10^{-3}$\\
$1.\,\allowbreak908\,9\times10^{-5}$%
\end{tabular}
&
\begin{tabular}
[c]{c}%
$16.\,\allowbreak83$\\
$186.88$%
\end{tabular}
&
\begin{tabular}
[c]{c}%
$\allowbreak8.\,\allowbreak45\,$\\
$25.\,\allowbreak53$%
\end{tabular}
$\allowbreak$ &
\begin{tabular}
[c]{c}%
$7$\\
$25$%
\end{tabular}
\\\hline
$^{197}$Au$^{2}$H & $X\;^{1}\Sigma^{+}\allowbreak$ & $\allowbreak
1.\,\allowbreak324\,5\times10^{-2}$ &
$1.\,\allowbreak761\,5\times10^{-5}$ &
$\allowbreak194.\,\allowbreak54$ & $35.\,\allowbreak51$ &
$\allowbreak 34$\\\hline $^{11}$B$^{35}$Cl & $A\;^{1}\Pi$ &
$1.\,\allowbreak340\,7\times10^{-2}$ &
$\allowbreak1.\,\allowbreak179\,2\times10^{-4}$ &
$\allowbreak75.\,\allowbreak 19$ & $\allowbreak28.\,\allowbreak88$
& $\allowbreak26$\\\hline
$^{9}$Be$^{1}$H & $%
\begin{array}
[c]{c}%
A\;^{2}\Pi\\
X\;^{2}\Sigma^{+}\allowbreak
\end{array}
$ & $%
\begin{array}
[c]{c}%
1.\,\allowbreak906\,4\times10^{-2}\\
\allowbreak1.\,\allowbreak724\,5\times10^{-2}%
\end{array}
$ & $%
\begin{array}
[c]{c}%
2.\,\allowbreak395\times10^{-4}\\
\allowbreak2.\,\allowbreak428\,8\times10^{-4}%
\end{array}
$ & $%
\begin{array}
[c]{c}%
52.\,\allowbreak76\\
\allowbreak52.\,39
\end{array}
\allowbreak$ & $%
\begin{array}
[c]{c}%
20.\,\allowbreak14\\
\allowbreak21.\,\allowbreak40
\end{array}
\allowbreak$ & $%
\begin{array}
[c]{c}%
18\\
\allowbreak19
\end{array}
\allowbreak$\\\hline
$\left(  ^{9}\text{Be}^{1}\text{H}\right)  \!^{+}$ & $%
\begin{array}
[c]{c}%
A\;^{2}\Pi\\
X\;^{2}\Sigma^{+}\allowbreak
\end{array}
\allowbreak$ & $%
\begin{array}
[c]{c}%
1.\,\allowbreak002\,6\times10^{-2}\\
\allowbreak.0\,\allowbreak179\,1
\end{array}
$ & $%
\begin{array}
[c]{c}%
2.\,\allowbreak574\,4\times10^{-5}\\
\allowbreak9.\,\allowbreak452\,2\times10^{-6}%
\end{array}
$ & $%
\begin{array}
[c]{c}%
160.\,\allowbreak92\\
265.58
\end{array}
$ & $%
\begin{array}
[c]{c}%
\allowbreak44.\,\allowbreak14\\
27.\,\allowbreak02
\end{array}
$ & $%
\begin{array}
[c]{c}%
42\\
26
\end{array}
$\\\hline
$\left(  ^{9}\text{Be}^{2}\text{H}\right)  \!^{+}$ & $%
\begin{array}
[c]{c}%
A\;^{2}\Pi\\
X\;^{2}\Sigma^{+}\allowbreak
\end{array}
$ & $%
\begin{array}
[c]{c}%
\allowbreak7.\,\allowbreak743\,5\times10^{-3}\\
1.\,\allowbreak326\,1\times10^{-2}%
\end{array}
$ & $%
\begin{array}
[c]{c}%
\allowbreak1.\,\allowbreak459\,3\times10^{-4}\\
\allowbreak3.\,\allowbreak641\,6\times10^{-5}%
\end{array}
$ & $%
\begin{array}
[c]{c}%
\allowbreak67.\,\allowbreak59\\
\allowbreak135.\allowbreak30
\end{array}
$ & $%
\begin{array}
[c]{c}%
\allowbreak36.\,\allowbreak29\\
33.\,\allowbreak90
\end{array}
$ & $%
\begin{array}
[c]{c}%
32\\
32
\end{array}
$\\\hline $^{9}\text{Be}^{16}$O & $B\;^{1}\Sigma^{+}\allowbreak$ &
$5.\,\allowbreak
650\,3\times10^{-3}\allowbreak\allowbreak\allowbreak$ &
$1.\,\allowbreak 969\,6\times10^{-7}$ & $\allowbreak1839.8$ &
$\allowbreak87.\,\allowbreak72$ & $\allowbreak87$\\\hline
$^{209}$Bi$_{2}$ & $%
\begin{tabular}
[c]{c}%
$B$\\
$X\;^{1}\Sigma_{g}^{+}\allowbreak$%
\end{tabular}
\ \ \ $ & $\
\begin{tabular}
[c]{c}%
$\allowbreak2.\,\allowbreak275\,92\times10^{-3}$\\
$\allowbreak1.\,\allowbreak868\,45\times10^{-3}$%
\end{tabular}
\ $ & $\
\begin{tabular}
[c]{c}%
$\allowbreak3.\,\allowbreak585\,21\times10^{-6}$\\
$\allowbreak1.\,\allowbreak343\,29\times10^{-5}$%
\end{tabular}
$ & $\
\begin{tabular}
[c]{c}%
$431.22$\\
$222.78$%
\end{tabular}
$ & $%
\begin{tabular}
[c]{c}%
$170.86$\\
$130.53$%
\end{tabular}
$ & $%
\begin{tabular}
[c]{c}%
$159$\\
$117$%
\end{tabular}
$\\\hline $^{209}$Bi$^{79}$Br & $A$ &
$3.\,\allowbreak929\,07\times10^{-3}$ &
$\allowbreak7.\,\allowbreak578\,54\times10^{-4}$ & $\allowbreak
29.\,\allowbreak66$ & $\allowbreak21.\,\allowbreak07$ &
$\allowbreak 18$\\\hline
$^{12}$C$_{2}$ & $%
\begin{array}
[c]{c}%
c\;^{1}\Pi_{g}\\
A\;^{3}\Pi_{g}%
\end{array}
$ & $%
\begin{array}
[c]{c}%
8.\,\allowbreak739\,2\times10^{-3}\\
9.\,\allowbreak193\,5\times10^{-3}%
\end{array}
\allowbreak$ & $%
\begin{array}
[c]{c}%
2.\,\allowbreak222\,1\times10^{-3}\\
\allowbreak2.\,\allowbreak833\,5\times10^{-4}%
\end{array}
$ & $%
\begin{array}
[c]{c}%
17.32\\
48.\,\allowbreak51
\end{array}
$ & $%
\begin{array}
[c]{c}%
11.80\\
\allowbreak27.\,\allowbreak42
\end{array}
$ & $%
\begin{array}
[c]{c}%
10\\
24
\end{array}
$\\\hline
$\left(  ^{35}\text{Cl}_{2}\right)  ^{+}$ & $%
\begin{array}
[c]{c}%
\left(  A\;^{2}\Pi\!\right)  \\
\left(  \!X\;^{2}\Pi\!\right)  \!\!^{a}%
\end{array}
\allowbreak$ & $%
\begin{array}
[c]{c}%
\allowbreak9.\,\allowbreak295\,8\times10^{-3}\\
7.\,\allowbreak312\,3\times10^{-3}%
\end{array}
\allowbreak\allowbreak\allowbreak$ & $%
\begin{array}
[c]{c}%
\allowbreak2.\,\allowbreak271\,5\times10^{-5}\\
6.\,\allowbreak728\times10^{-5}%
\end{array}
$ & $%
\begin{array}
[c]{c}%
171.\,\allowbreak32\\
\allowbreak99.\,\allowbreak54
\end{array}
$ & $%
\begin{array}
[c]{c}%
47.\,\allowbreak53\\
46.\,\allowbreak36
\end{array}
$ & $%
\begin{array}
[c]{c}%
45\\
42
\end{array}
$\\\hline $^{35}$Cl$^{19}$F & $A\;^{3}\Pi_{0^{+}}$\  &
$\allowbreak6.\,\allowbreak 688\,2\times10^{-3}$ &
$1.\,\allowbreak206\,7\times10^{-3}$ & $\allowbreak
23.\,\allowbreak51$ & $16.13$ & $14$\\\hline $^{12}$C$^{16}$O &
$d\;^{3}\Pi_{i}$\  & $6.\,\allowbreak700\,7\times10^{-3}$ &
$9.\,\allowbreak887\,6\times10^{-5}$ & $82.\,\allowbreak112$ &
$\allowbreak 43.\,\allowbreak34$ & $39$\\\hline $\left(
^{12}\text{C}^{16}\text{O}\right)  \!^{+}$ & $X\;^{2}\Sigma
^{+}\allowbreak$ & $6.\,\allowbreak848\,4\times10^{-3}$ &
$3.\,\allowbreak 161\,4\times10^{-7}$ & $1452.\,\allowbreak2$ &
$\allowbreak72.\,\allowbreak26$ & $72$\\\hline
$^{133}$Cs$_{2}$ & $%
\begin{array}
[c]{c}%
B\;\left(  ^{1}\Pi_{u}\!\right)  \\
X\;^{1}\Sigma_{g}^{+}\allowbreak
\end{array}
$ & $%
\begin{array}
[c]{c}%
2.\,\allowbreak278\,4\times10^{-3}\\
1.\,\allowbreak906\,4\times10^{-3}%
\end{array}
$ & $%
\begin{array}
[c]{c}%
5.\,\allowbreak495\,2\times10^{-6}\\
3.\,\allowbreak912\,8\times10^{-6}%
\end{array}
$ & $%
\begin{array}
[c]{c}%
348.\,\allowbreak31\\
\allowbreak412.\,\allowbreak77
\end{array}
$ & $%
\begin{array}
[c]{c}%
156.\,\allowbreak24\\
\allowbreak186.\,\allowbreak09
\end{array}
$ & $%
\begin{array}
[c]{c}%
143\\
171
\end{array}
$\\\hline $^{63}$Cu$^{2}$H & $A\;^{1}\Sigma^{+}\allowbreak$ &
$\allowbreak 1.\,\allowbreak702\,2\times10^{-2}$ &
$\allowbreak3.\,\allowbreak 379\,6\times10^{-4}$ &
$\allowbreak44.\,\allowbreak41$ & $\allowbreak 20.\,\allowbreak01$
& $\allowbreak18$\\\hline $^{1}$H$_{2}$ &
$e\;^{3}\Sigma_{u}^{+}\allowbreak$ & $\allowbreak
2.\,\allowbreak996\,6\times10^{-2}$ & $\allowbreak1.\,\allowbreak
971\,9\times10^{-4}$ & $\allowbreak58.\,\allowbreak15$ &
$\allowbreak 14.\,\allowbreak65$ & $14$\\\hline
$^{202}$Hg$^{81}$Br & $X\;\left(  ^{3}\Sigma\right) $ &
$\allowbreak 5.\,\allowbreak209\,5\times10^{-3}$ &
$4.\,\allowbreak833\,5\times10^{-5}$ &
$\allowbreak117.\,\allowbreak44$ & $59.50$ & $54$\\\hline
Hg$^{1}$H & $%
\begin{array}
[c]{c}%
A\;^{2}\Pi_{3/2}\\
X\;^{2}\Sigma^{+}\allowbreak
\end{array}
$ & $%
\begin{array}
[c]{c}%
2.\,\allowbreak024\,8\times10^{-2}\\
5.\,\allowbreak984\,5\times10^{-2}%
\end{array}
$ & $%
\begin{array}
[c]{c}%
2.\,\allowbreak080\,4\times10^{-4}\\
2.\,\allowbreak126\,8\times10^{-3}%
\end{array}
$ & $%
\begin{array}
[c]{c}%
56.\,\allowbreak61\\
\allowbreak17.\,\allowbreak70
\end{array}
$ & $%
\begin{array}
[c]{c}%
\allowbreak19.\,\allowbreak81\\
6.\,\allowbreak20
\end{array}
$ & $%
\begin{array}
[c]{c}%
18\\
5
\end{array}
$\\\hline $\left(  \text{Hg}^{1}\text{H}\right)  ^{+}$ &
$A\;^{1}\Sigma^{+}\allowbreak$ &
$2.\,\allowbreak656\,4\times10^{-2}$ &
$4.\,\allowbreak721\,2\times10^{-3}$ &
$\allowbreak11.\,\allowbreak88$ & $7.\,\allowbreak02$ &
$6$\\\hline $\left(  \text{Hg}^{2}\text{H}\right)  ^{+}$ &
$A\;^{1}\Sigma^{+}\allowbreak$ &
$1.\,\allowbreak886\,5\times10^{-2}$ &
$2.\,\allowbreak375\,9\times10^{-3}$ &
$\allowbreak16.\,\allowbreak75$ & $10.09$ & $8$\\\hline
$^{127}$I$_{2}$ & $%
\begin{array}
[c]{c}%
G\;\left(  ^{1}\Sigma_{u}^{+}\right)  \\
X\;^{1}\Sigma_{g}^{+}\allowbreak
\end{array}
$ & $%
\begin{array}
[c]{c}%
3.\,\allowbreak603\,9\times10^{-3}\\
2.\,\allowbreak855\,5\times10^{-3}%
\end{array}
$ & $%
\begin{array}
[c]{c}%
2.\,\allowbreak119\,9\times10^{-5}\\
4.\,\allowbreak171\,1\times10^{-6}%
\end{array}
$ & $%
\begin{array}
[c]{c}%
177.\,\allowbreak34\\
399.\,\allowbreak79
\end{array}
$ & $%
\begin{array}
[c]{c}%
88.36\\
143.\,\allowbreak3
\end{array}
$ & $%
\begin{array}
[c]{c}%
80\\
134
\end{array}
$\\\hline $^{127}$I$^{35}$Cl & $A\;^{3}\Pi_{1}\allowbreak$ &
$9.\,\allowbreak 284\,69\times10^{-3}$ &
$\allowbreak1.\,\allowbreak605\,15\times10^{-4}$ &
$64.\,\allowbreak45$ & $32.74$ & $29$\\\hline $^{115}$In$^{1}$H &
$A\;^{1}\Sigma^{+}$ & $4.\,\allowbreak175\,81\times 10^{-2}$ &
$4.\,\allowbreak696\,93\times10^{-3}$ & $\allowbreak
11.\,\allowbreak91$ & $\allowbreak6.\,\allowbreak107$ &
$\allowbreak5$\\\hline In$^{16}$O & $\left(  X\;^{2}\Sigma\right)
^{a}$ & $5.\,\allowbreak 276\,71\times10^{-3}$ &
$4.\,\allowbreak053\,54\times10^{-4}$ & $\allowbreak
40.\,\allowbreak55$ & $\allowbreak27.\,\allowbreak08$ &
$24$\\\hline $^{7}$Li$_{2}$ &
\begin{tabular}
[c]{c}%
$B\;^{1}\Pi_{u}$\\
$X\;^{1}\Sigma_{g}^{+}\allowbreak$%
\end{tabular}
&
\begin{tabular}
[c]{c}%
$1.\,\allowbreak017\,46\times10^{-2}$\\
$7.\,\allowbreak375\,58\times10^{-3}$%
\end{tabular}
$\allowbreak$ &
\begin{tabular}
[c]{c}%
$2.\,\allowbreak361\,97\times10^{-4}$\\
$1.\,\allowbreak650\,40\times10^{-5}$%
\end{tabular}
$\allowbreak$ &
\begin{tabular}
[c]{c}%
$53.\,\allowbreak13$\\
$200.98$%
\end{tabular}
&
\begin{tabular}
[c]{c}%
$28.\,\allowbreak06$\\
$59.14$%
\end{tabular}
$\allowbreak$ &
\begin{tabular}
[c]{c}%
$\allowbreak25$\\
$56$%
\end{tabular}
\\\hline
\end{tabular}

\end{table}

\begin{table}
\setcounter{table}{1} \caption{continues}
\begin{tabular}
[c]{|c|c|c|c|c|c|c|}\hline molecule & state, line & $x_{e}$ &
$\left|  y_{e}\right| $ & $p$ & $j$ & $\left[  n_{-}\right]
$\\\hline $^{\left(  24\right)  }$Mg$^{1}$H &
$X\;^{2}\Sigma^{+}\allowbreak$ &
$\allowbreak2.\,\allowbreak106\,04\times10^{-2}$ &
$1.\,\allowbreak 002\,87\times10^{-4}$ &
$\allowbreak81.\,\allowbreak53$ & $\allowbreak 21.\,\allowbreak00$
& $\allowbreak20$\\\hline $\left(  \!^{\left(  24\right)
}\text{Mg}^{1}\text{H\negthinspace}\right)
\!^{+}$ & $%
\begin{array}
[c]{c}%
A\;^{1}\Sigma^{+}\allowbreak\\
X\;^{2}\Sigma^{+}\allowbreak
\end{array}
$ & $%
\begin{array}
[c]{c}%
6.\,\allowbreak003\,35\times10^{-3}\\
1.\,\allowbreak781\,40\times10^{-2}%
\end{array}
$ &
\begin{tabular}
[c]{c}%
$\allowbreak3.\,\allowbreak178\,25\times10^{-4}$\\
$3.\,\allowbreak008\,32\times10^{-4}$%
\end{tabular}
&
\begin{tabular}
[c]{c}%
$45.80$\\
$47.08$%
\end{tabular}
&
\begin{tabular}
[c]{c}%
$29.33$\\
$20.05$%
\end{tabular}
&
\begin{tabular}
[c]{c}%
$26$\\
$18$%
\end{tabular}
\\\hline
$\left(  \!^{\left(  24\right)  }\!\text{Mg}^{2}\text{H}\right)  \!^{+}$ & $%
\begin{array}
[c]{c}%
A\;^{1}\Sigma^{+}\allowbreak\\
X\;^{2}\Sigma^{+}\allowbreak
\end{array}
$ &
\begin{tabular}
[c]{c}%
$\allowbreak4.\,\allowbreak247\,25\times10^{-3}$\\
$1.\,\allowbreak328\,88\times10^{-2}$%
\end{tabular}
&
\begin{tabular}
[c]{c}%
$1.\,\allowbreak432\,07\times10^{-4}$\\
$1.\,\allowbreak361\,49\times10^{-4}$%
\end{tabular}
&
\begin{tabular}
[c]{c}%
$68.23$\\
$69.\,\allowbreak98$%
\end{tabular}
&
\begin{tabular}
[c]{c}%
$\allowbreak43.\,\allowbreak46$\\
$28.25$%
\end{tabular}
&
\begin{tabular}
[c]{c}%
$38$\\
$26$%
\end{tabular}
\\\hline
$^{14}$N$_{2}$ &
\begin{tabular}
[c]{c}%
$C\;^{3}\Pi_{u}$\\
$A\;^{3}\Sigma_{u}^{+}\allowbreak$\\
$a\;^{1}\Pi_{g}$%
\end{tabular}
&
\begin{tabular}
[c]{c}%
$\allowbreak8.\,\allowbreak392\,71\times10^{-3}$\\
$\allowbreak9.\,\allowbreak511\,97\times10^{-3}$\\
$7.\,\allowbreak559\,65\times10^{-3}$%
\end{tabular}
&
\begin{tabular}
[c]{c}%
$1.\,\allowbreak056\,46\times10^{-3}$\\
$1.\,\allowbreak711\,89\times10^{-5}$\\
$2.\,\allowbreak062\,04\times10^{-4}$%
\end{tabular}
&
\begin{tabular}
[c]{c}%
$\allowbreak25.\,\allowbreak12$\\
$197.\,\allowbreak34$\\
$56.\,\allowbreak86$%
\end{tabular}
&
\begin{tabular}
[c]{c}%
$16.\,\allowbreak62$\\
$\allowbreak47.\,\allowbreak80$\\
$\allowbreak32.\,\allowbreak62$%
\end{tabular}
&
\begin{tabular}
[c]{c}%
$14$\\
$46$\\
$29$%
\end{tabular}
\\\hline
$\left(  ^{14}\text{N}_{2}\right)  ^{+}$ &
\begin{tabular}
[c]{c}%
$B\;^{2}\Sigma_{u}^{+}\allowbreak$\\
$X\;^{2}\Sigma_{g}^{+}\allowbreak$%
\end{tabular}
&
\begin{tabular}
[c]{c}%
$9.\,\allowbreak583\,28\times10^{-3}$\\
$7.\,\allowbreak310\,65\times10^{-3}$%
\end{tabular}
&
\begin{tabular}
[c]{c}%
$2.\,\allowbreak221\,22\times10^{-4}$\\
$1.\,\allowbreak812\,26\times10^{-5}$%
\end{tabular}
&
\begin{tabular}
[c]{c}%
$\allowbreak54.\,\allowbreak79$\\
$191.80$%
\end{tabular}
&
\begin{tabular}
[c]{c}%
$\allowbreak29.\,\allowbreak28$\\
$58.91$%
\end{tabular}
&
\begin{tabular}
[c]{c}%
$26$\\
$56$%
\end{tabular}
\\\hline
$^{23}$Na$_{2}$ & $%
\begin{array}
[c]{c}%
B\;^{1}\Pi_{u}\\
X\;^{1}\Sigma_{g}^{+}\allowbreak
\end{array}
$ & $%
\begin{array}
[c]{c}%
5.\,\allowbreak091\,69\times10^{-3}\\
4.\,\allowbreak559\,44\times10^{-3}%
\end{array}
$ & $%
\begin{array}
[c]{c}%
7.\,\allowbreak561\,19\times10^{-5}\\
\allowbreak1.\,\allowbreak695\,66\times10^{-5}%
\end{array}
$ & $%
\begin{array}
[c]{c}%
\allowbreak93.\,\allowbreak90\\
198.\,\allowbreak28
\end{array}
$ & $%
\begin{array}
[c]{c}%
\allowbreak52.\,\allowbreak31\\
\allowbreak82.\,\allowbreak36
\end{array}
$ & $%
\begin{array}
[c]{c}%
47\\
76
\end{array}
$\\\hline $^{23}$Na$^{1}$H & $A\;^{1}\Sigma^{+}$ &
$\allowbreak1.\,\allowbreak 741\,79\times10^{-2}$ &
$\allowbreak6.\,\allowbreak342\,56\times10^{-4}$ &
$\allowbreak32.\,\allowbreak42$ & $\allowbreak16.\,\allowbreak63$
& $\allowbreak15$\\\hline Na$^{23}$K &
\begin{tabular}
[c]{c}%
$D\;^{1}\Pi$\\
$A\;^{1}\Sigma^{+}$%
\end{tabular}
&
\begin{tabular}
[c]{c}%
$4.\,\allowbreak258\,94\times10^{-3}$\\
$\allowbreak1.\,\allowbreak092\,05\times10^{-3}$%
\end{tabular}
&
\begin{tabular}
[c]{c}%
$\allowbreak9.\,\allowbreak905\,09\times10^{-5}$\\
$4.\,\allowbreak871\,63\times10^{-5}$%
\end{tabular}
&
\begin{tabular}
[c]{c}%
$82.04$\\
$116.\,\allowbreak98$%
\end{tabular}
&
\begin{tabular}
[c]{c}%
$50.14$\\
$83.95$%
\end{tabular}
&
\begin{tabular}
[c]{c}%
$44$\\
$75$%
\end{tabular}
\\\hline
$^{14}$N$^{16}$O & $%
\begin{array}
[c]{c}%
\;A\;^{2}\Sigma^{+}\\
X\;\;^{2}\Pi_{3/2}\\
X\;\;^{2}\Pi_{1/2}%
\end{array}
$ & $%
\begin{array}
[c]{c}%
\allowbreak6.\,\allowbreak106\,36\times10^{-3}\\
7.\,\allowbreak338\,42\times10^{-3}\\
7.\,\allowbreak337\,07\times10^{-3}%
\end{array}
$ & $%
\begin{array}
[c]{c}%
\allowbreak1.\,\allowbreak180\,79\times10^{-4}\\
\allowbreak6.\,\allowbreak303\,58\times10^{-7}\\
\allowbreak6.\,\allowbreak302\,42\times10^{-7}%
\end{array}
$ & $%
\begin{array}
[c]{c}%
\allowbreak75.\,\allowbreak14\\
1028.40\\
\allowbreak1028.\,\allowbreak49
\end{array}
$ & $%
\begin{array}
[c]{c}%
42.\,\allowbreak35\\
67.24\\
67.25
\end{array}
$ & $%
\begin{array}
[c]{c}%
38\\
67\\
67
\end{array}
$\\\hline
$^{16}$O$_{2}$ & $%
\begin{array}
[c]{c}%
B\;^{3}\Sigma_{u}^{-}\\
b\;^{1}\Sigma_{g}^{+}%
\end{array}
$ & $%
\begin{array}
[c]{c}%
1.\,\allowbreak142\,56\times10^{-2}\\
\allowbreak9.\,\allowbreak736\,95\times10^{-3}%
\end{array}
$ & $%
\begin{array}
[c]{c}%
5.\,\allowbreak358\,67\times10^{-4}\\
\allowbreak7.\,\allowbreak503\,38\times10^{-6}%
\end{array}
$ & $%
\begin{array}
[c]{c}%
\allowbreak35.\,\allowbreak27\\
\allowbreak298.\,\allowbreak08
\end{array}
$ & $%
\begin{array}
[c]{c}%
20.45\\
48.95
\end{array}
$ & $%
\begin{array}
[c]{c}%
18\\
48
\end{array}
$\\\hline $^{16}$O$^{1}$H & $A\;^{2}\Sigma^{+}$ &
$\allowbreak2.\,\allowbreak 984\,75\times10^{-2}$ &
$\allowbreak2.\,\allowbreak034\,27\times10^{-4}$ & $57.25$ &
$14.66$ & $14$\\\hline $^{31}$P$_{2}$ & $X\;^{1}\Sigma_{g}^{+}$ &
$\allowbreak3.\,\allowbreak 592\,89\times10^{-3}$ &
$6.\,\allowbreak829\,57\times10^{-6}$ & $312.43$ & $113.18$ &
$106$\\\hline Pb$^{79}$Br & $A\;\left(  ^{2}\Sigma\right) $ &
$\allowbreak2.\,\allowbreak 622\,95\times10^{-3}$ &
$\allowbreak1.\,\allowbreak836\,07\times10^{-4}$ & $60.26$ &
$42.11$ & $37$\\\hline PbSe & $D$ &
$2.\,\allowbreak783\,61\times10^{-3}$ & $2.\,\allowbreak
100\,84\times10^{-5}$ & $178.14$ & $98.41$ & $88$\\\hline
$^{85}$Rb$_{2}$ & $%
\begin{array}
[c]{c}%
C\\
X\;^{1}\Sigma_{g}^{+}%
\end{array}
$ & $%
\begin{array}
[c]{c}%
1.\,\allowbreak843\,15\times10^{-3}\\
1.\,\allowbreak675\,98\times10^{-2}%
\end{array}
$ & $%
\begin{array}
[c]{c}%
\\
1.\,\allowbreak449\,02\times10^{-5}%
\end{array}
\allowbreak$ & $%
\begin{array}
[c]{c}%
136.\,\allowbreak80\\
214.50
\end{array}
$ & $%
\begin{array}
[c]{c}%
\allowbreak90.\,\allowbreak04\\
28.60
\end{array}
$ & $%
\begin{array}
[c]{c}%
80\\
28
\end{array}
$\\\hline $^{28}$Si$^{32}$S &
\begin{tabular}
[c]{c}%
$E$\\
$D\;^{1}\Pi$%
\end{tabular}
&
\begin{tabular}
[c]{c}%
$3.\,\allowbreak469\,64\times10^{-3}$\\
$\allowbreak4.\,\allowbreak648\,44\times10^{-3}$%
\end{tabular}
&
\begin{tabular}
[c]{c}%
$\allowbreak8.\,\allowbreak153\,66\times10^{-5}$\\
$\allowbreak8.\,\allowbreak789\,06\times10^{-5}$%
\end{tabular}
&
\begin{tabular}
[c]{c}%
$90.43$\\
$\allowbreak87.\,\allowbreak09$%
\end{tabular}
&
\begin{tabular}
[c]{c}%
$56.77$\\
$51.15$%
\end{tabular}
&
\begin{tabular}
[c]{c}%
$50$\\
$45$%
\end{tabular}
\\\hline
$^{28}$SiSe & $E$ &
$\allowbreak6.\,\allowbreak314\,77\times10^{-3}$ &
$\allowbreak1.\,\allowbreak036\,27\times10^{-4}$ & $80.21$ &
$43.69$ & $39$\\\hline Sn$^{16}$O & $D\;^{1}\Sigma^{+}$ &
$\allowbreak5.\,\allowbreak286\,65\times 10^{-3}$ &
$\allowbreak2.\,\allowbreak317\,20\times10^{-4}$ & $53.64$ &
$34.22$ & $30$\\\hline SnS & $E$ &
$\allowbreak3.\,\allowbreak694\,29\times10^{-3}$ & $\allowbreak
4.\,\allowbreak067\,11\times10^{-5}$ & $128.03$ & $71.77$ &
$64$\\\hline SnSe & $E$ &
$\allowbreak3.\,\allowbreak916\,58\times10^{-3}$ & $\allowbreak
8.\,\allowbreak138\,35\times10^{-6}$ & $286.\,\allowbreak21$ &
$103.74$ & $97$\\\hline
SnTe & $%
\begin{array}
[c]{c}%
I\\
B
\end{array}
$ & $%
\begin{array}
[c]{c}%
\allowbreak5.\,\allowbreak441\,88\times10^{-3}\\
\allowbreak6.\,\allowbreak643\,51\times10^{-3}%
\end{array}
$ & $%
\begin{array}
[c]{c}%
1.\,\allowbreak306\,05\times10^{-5}\\
\allowbreak5.\,\allowbreak644\,81\times10^{-5}%
\end{array}
$ & $%
\begin{array}
[c]{c}%
225.93\\
108.68
\end{array}
$ & $%
\begin{array}
[c]{c}%
76.64\\
50.87
\end{array}
$ & $%
\begin{array}
[c]{c}%
72\\
46
\end{array}
$\\\hline Sr$^{19}$F & $B\;^{2}\Sigma$ &
$\allowbreak3.\,\allowbreak804\,46\times 10^{-3}$ &
$1.\,\allowbreak431\,79\times10^{-5}$ & $215.78$ & $94.81$ &
$87$\\\hline Tl$^{81}$Br & $A\;\left(  ^{3}\Pi_{0}^{+}\right)
^{b}$ & $4.\,\allowbreak 754\,43\times10^{-2}$ &
$\allowbreak2.\,\allowbreak031\,02\times10^{-3}$ & $18.12$ &
$7.29$ & $6$\\\hline $\left(  \text{Zn}^{1}\text{H}\right)  ^{+}$
& $X\;^{1}\Sigma^{+}$ &
$\allowbreak2.\,\allowbreak035\,49\times10^{-2}$ &
$\allowbreak1.\,\allowbreak 043\,84\times10^{-4}$ &
$79.\,\allowbreak92$ & $\allowbreak21.\,\allowbreak53$ &
$\allowbreak20$\\\hline
\end{tabular}
\end{table}

There are six electronic states $A\;^{1}\Sigma^{+}\;$of the diatomic
molecules $^{133}$Cs$^{1}$H,\ $^{\left( 39\right) }$K$^{2}$H,\ $^{7}$%
Li$^{1}$H,\ $^{7}$Li$^{2}$H,\ $^{23}$Na$^{1}$H and Rb$^{1}$H in $\ $%
\cite{herz:dia}, for which the values of both molecular constants
$x_{e}$ and $y_{e}$ are negative. The model can accommodate these
cases and the values of the parameters $p$ and $j^{\prime}$,
calculated from (\ref{lanni}) and (\ref{lappiz}), are given in
Table 3. The number of vibrational levels is determined by
$[j^{\prime}]$. As we have pointed out the case $x_{e}<0$ is not a
deformation of the Morse potential. Thus, it does not behave like
(deformed) oscillator models do. In spite of this, our model can
handle it.\\

\begin{table}
\caption{Parameters of the model for negative molecular constants
$x_e$ and $y_e$}
\begin{tabular}{|c|c|c|c|c|c|c|}
\hline
molecule & state & $\left| x_{e}\right| $ & $\left| y_{e}\right| $ & $p$ & $%
j^{\prime }$ & $j$ \\ \hline $^{133}$Cs$^{1}$H & $A\;^{1}\Sigma
^{+}$ & $\allowbreak 2.\,\allowbreak
794\,12\times 10^{-2}$ & $1.\,\allowbreak 715\,69\times 10^{-3}$ & $%
\allowbreak 19.\,\allowbreak 71$ & $10.02$ & $20.\,\allowbreak 94$
\\ \hline $^{\left( 39\right) }$K$^{2}$H & $A\;^{1}\Sigma ^{+}$ &
$\allowbreak 1.\,\allowbreak 133\,22\times 10^{-2}$ & $\allowbreak
2.\,\allowbreak 745\,61\times 10^{-4}$ & $49.\,\allowbreak 28$ &
$25.65$ & $51.75$ \\ \hline
$^{7}$Li$^{1}$H & $A\;^{1}\Sigma ^{+}$ & $\allowbreak 0.123\,487$ & $%
1.\,\allowbreak 785\,27\times 10^{-2}$ & $\allowbreak
6.\,\allowbreak 111$ & $\allowbreak 2.\,\allowbreak 324$ & $7.275$
\\ \hline $^{7}$Li$^{2}$H & $A\;^{1}\Sigma ^{+}$ & $\allowbreak
7.\,\allowbreak 739\,98\times 10^{-2}$ & $\allowbreak
6.\,\allowbreak 520\,91\times 10^{-3}$ & $10.11$ & $\allowbreak
4.\,\allowbreak 084\,$ & $11.80$ \\ \hline $^{23}$Na$^{1}$H &
$A\;^{1}\Sigma ^{+}$ & $\allowbreak \allowbreak 1.\,\allowbreak
741\,79\times 10^{-2}$ & $\allowbreak 6.\,\allowbreak
342\,56\times 10^{-4}$ & $32.\,\allowbreak 42$ & $16.63$ & $34.30$
\\ \hline Rb$^{1}$H & $A\;^{1}\Sigma ^{+}$ & $1.\,\allowbreak
676\,21\times 10^{-2}$ &
$\allowbreak \allowbreak 6.\,\allowbreak 909\,24\times 10^{-4}$ & $%
\allowbreak 31.\,\allowbreak 06$ & $\allowbreak 16.\,\allowbreak 441$ & $%
32.\,\allowbreak 35$ \\ \hline
\end{tabular}
\end{table}

\section{Conclusions and Outlook}

In this paper we have considered a $q$-deformation of a general
Hamiltonian constructed from the raising and lowering generators
of the quantum group deformation of the algebra $su(2)$ and have
obtained an expression for the eigenvalues of this Hamiltonian
which can be interpreted as the phenomenological Dunham's
expansion of the vibrational energies of diatomic molecules. The
expansion is truncated to the cubic term mainly for practical
reasons as the experimental results suggest that the contribution
of terms of higher order is often negligibly small and seldom
used. We have formulated a model of the anharmonic vibrations of
diatomic molecules which in different approximations leads to
Morse or Dunham's results. The parameters of the model are
obtained in terms of the well-known experimental molecular
constants and this gives a clear test for the applicability of the
model and its restrictions. The model can accommodate both the
positive and the negative values of the anharmonic constant
$y_{e}$. Using all data available in \cite{herz:dia}, we have
tested the model in a global fashion for a large set of diatomic
molecules and conclude that it fits well with the experimental
data for all states, except for about 30 electronic states for
which the values of $x_{e}$ are not much larger than the values of
$y_{e}$, as required by the model. In these cases the model could
be expanded by adding to the Hamiltonian terms corresponding to
the rotational energies in order to obtain a $q$-deformed version
corresponding to the general Dunham expansion (\ref{gdunham}).
Next, in order to test these ideas further, intensity data should
be studied. This would involve a procedure where a consistent
$q$-deformed dipole operator is defined and the wave functions for
each molecule evaluated within our approximation. Finally, a more
ambitious project would be to extend these studies to polyatomic
molecules, following, for example, the methodology of reference
\cite{Fra}.

\noindent {\large\bf Acknowledgements:} ~~The authors would like
to thank Dr. L.L. Boyle for very helpful discussions.


\end{document}